

              \def\date{October 1993}

\magnification=\magstep1
\baselineskip=.65cm

\hsize=13.7truecm
\voffset=.800truecm
\hoffset=0.50truecm

\font\caps=cmssbx10 at 14.4 true pt
\font\tilsev=cmsy7
\font\tilfive=cmsy5
\def\lap{\hbox{\hbox{$<$} \kern-1.0em\raise-1.1mm \hbox{\tilsev \char'030}}}
\def\gap{\hbox{\hbox{$>$} \kern-1.2em\raise-1.1mm \hbox{\tilsev \char'030}}}

{\baselineskip=.45cm\rightline{FUB-HEP 16/93}
                   \rightline{\date}        }

\bigskip
\bigskip
\centerline{\caps The crossover from first to second-order
finite-size scaling:}
\centerline{\caps a numerical study}

\vskip0.5truein
\centerline{\bf Christian Borgs and P.E.L. Rakow}
\centerline{\it Institut f\"ur Theoretische Physik,
Freie Universit\"at Berlin,}
\centerline{\it Arnimallee 14, D-14195 Berlin, Germany}
\vskip0.5truein
\centerline{\bf Stefan Kappler}
\centerline{\it Institut f\"ur Physik,
Universit\"at Mainz,}
\centerline{\it Staudinger Weg 7, D-55128 Mainz, Germany}
\vskip0.5truein
\noindent{\bf Abstract:} We consider a particular case
 of the two dimensional Blume-Emery-Griffiths model to study the
 finite-size scaling
 for a field driven first-order phase transition with two
 coexisting phases not related by a symmetry. For low temperatures
 we verify the asymptotic (large volume) predictions of the rigorous
  theory
 of Borgs and Koteck\'y, including the predictions concerning
 the so-called equal-weight versus equal-height controversy.
 Near the critical temperature we show that all data fit onto a
 unique curve, even when the correlation length $\xi$ becomes
 comparable to or larger then the size of the system,
 provided the linear dimension $L$ of the system is rescaled by $\xi$.

\vfill\eject

\noindent{\bf 1. Introduction}
\medskip

 First-order phase transitions play an important role in physics.
 Examples range from well known transitions in solid
 state physics (like ordinary melting) to phase transitions in
 the early universe.
 For many models describing such a transition,
 exact solutions are not available, and systematic expansions
 such as ``weak coupling" or ``high temperature series"
 are of limited value. In this situation, the numerical simulation
 of physically interesting models has gained more and more popularity.
 With the increasing speed of computers, the statistical
 errors of Monte Carlo simulations have been drastically reduced, but
 a systematic limitation of this method is the finite size of
 the simulated system.
 A clear understanding of finite-size effects is
 therefore more and more important in order to interpret
 the resulting numerical data.

  The theory of finite-size scaling (FSS) for second-order
 phase transitions has a rich and longstanding
 literature, starting with the pioneering work of Fisher
 [1] and Fisher and Barber [2]. In the last ten years,
 the theory of FSS has been extended to  first-order transitions.
 For systems with two coexisting phases, three main approaches
 can be distinguished in the literature: the phenomenological renormalisation
 group approach to FSS of Berker/Fisher [3] and Bl\"ote/Nightingale [4],
 the thermodynamic fluctuation theory for the order parameter distribution
 $P_V(s)$ of Binder and co-workers [5,6,7] and the transfer matrix method
 of Privman and Fisher [8], see Refs.~[9] and [10] for reviews.

 Recently, the theory of FSS near first order transitions has been
 put on a rigorous basis [11,12] within the context of the
 Pirogov-Sinai theory [13]
 which applies whenever it is possible to describe the
 configurations of the system in terms of energetically unfavourable
 contours. Typical models to which these methods can be
 applied are field-driven transitions in low temperature spin
 systems with discrete spins, temperature-driven
 transitions for Potts models, lattice field
 theories with double well potential, Higgs/confinement transitions
 in large N lattice gauge theories, and certain $2d$
 continuum field theories, such as $P(\phi)_2$ field theories in
 the broken phase and supersymmetric Wess-Zumino models.

 In the context of the field-driven transitions with two
 coexisting phases considered in this paper, the main
 result of Ref.~[11] can be summarized
 by a formula of the form
$$
  Z(V,h)=
  \left[
  e^{-\beta f_+(\beta,h)V} + e^{-\beta f_-(\beta,h)V}
  \right]
  (1+O(Ve^{-L/L_0}))
 \eqno(1.1)
$$
 for the partition function in a cubic volume $V=L^d$ with periodic
 boundary conditions [14]. Here $L_0$ is a constant
 of the order of the infinite volume correlation length, and $f_m(\beta,h)$
 is some sort of metastable free energy for the phase $m$. It is equal
 to the free energy $f(\beta,h)$ if $m$ is stable, strictly
 larger then $f(\beta,h)$ if $m$ is unstable, and may be chosen as a smooth
 function of $h$. The finite size scaling of the partition
 function and its derivatives is obtained by expanding
 $f_m(\beta,h)$ around the transition point $h_t$.

  In a recent  paper [16], we have analysed a controversy
 stemming from the work of Binder and co-workers: within the
 phenomenological fluctuation theory for the order parameter
 distribution $P_V(s)$, different formulae for the
 FSS of the order parameter
 are obtained assuming different ans\"atze [6,7] for the
 relative normalisation of the two Gaussians.
 If the normalisation is chosen in such
 a way that both Gaussians have {\it equal weight} at the infinite
 volume transition point $h_t$, the resulting formula for the magnetisation
 in a cube with periodic boundary conditions
 leads to the prediction [6] that the shift of the susceptibility
 maximum with respect to $h_t$ is proportional to $L^{-2d}$,
 while a normalisation with {\it equal peak height} at $h_t$ leads
 to the prediction [7] that this
 shift is proportional to
 $L^{-d}$.

  To some extent, this controversy had already been resolved in
  Ref.~[11], where it had been shown that the predictions of the
  {\it equal weight} prescription agree with those following from
 equation~(1.1) above.  Unfortunately, equation~(1.1) is
 only {\it proven} for very low temperatures (for Ising type systems
 in $d=2$, this restricts the temperature to about half $T_c$), while most
 numerical simulations are performed near $T_c$. It therefore
 seemed desirable to test the predictions following from (1.1)
 in an actual numerical
 simulation in order to definitely settle the controversy between [6]
 and [7].

  This was done in [16]. We considered
  a spin model with spins $s_x\in\{-1,0,1\}$
  and Hamiltonian
$$
   H={1\over 2}\sum_{<xy>} |\sigma(s_x)-\sigma(s_y)|^2 - h\sum_x s_x
  \;,
   \eqno(1.2{\rm a})
$$
  where the first sum goes over the set of $dL^d$ nearest neighbours
   in $V$ and
$$
   \sigma(s_x)=\cases{
   -1&if $s_x=-1$\cr
   +1&if $s_x=0,+1\;.$\cr}
   \eqno(1.2{\rm b})
$$
  It is defined in such a way that boundaries
  between $-1$ and $0$ or $+1$ are energetically unfavourable, while
  boundaries between $0$ and $+1$ do not cost energy. As a consequence,
  the low temperature phases of the model consist
  of a phase ``minus" corresponding to spin configurations where most
  spins $s_x$ are equal to $-1$ with small islands of $0$ or $+1$,
  and a phase ``plus"
  corresponding to spin configurations which are a mixture of
  $0$ and $+1$ with small islands of $-1$. The
  phase minus is stable as long as $h$ is smaller
  then a certain critical value $h_t=h_t(\beta)$,
  while the plus phase is stable for $h>h_t$.
  Because in the ``plus" phase spins can be flipped from $0$ to
  $+1$ without creating any energetically unfavourable boundaries
  we expect the magnetic susceptibility of this phase to be
  larger than that of the ``minus" phase. A large susceptibility
  difference between the phases is desirable because it leads to
  a clear  difference  between the predictions of the
  equal-weight and equal-height theories of first order finite
  size effects.

   This model is a special case of the Blume-Emery-Griffiths (BEG)
  model [17], which is the most general model with a three-valued spin
   and nearest neighbour interactions. The general BEG model has
  a complicated phase structure, with up to three coexisting phases.
  Here we have chosen the coupling constants in such a way
   that the model has two coexisting phases at the phase transition
   point $h_t$, with a large susceptibility difference.

   In [16], we simulated the model (1.2) in $d=2$ and determined
  the shift of the susceptibility maximum with respect to $h_t$.
  For $\beta=0.5$, and lattices with linear dimensions $L$
  from $4$ to $30$, we found overwhelming evidence for the
  equal weight prescription, as to be expected from the rigorous
  results of Ref.~[11]. We summarize the main results of Ref.~[16],
  as well as certain extensions (in particular to $\beta=0.45$
  and $\beta=0.47$) and refinements in Section 3.

  In [16], we did not analyse, however, other finite-size scaling predictions
  following from equation~(1.1), such as the volume dependence of the
  susceptibility-maximum, or the volume dependence of the susceptibility
  at the transition point $h_t$. In view of the results of
  Billoire and co-workers [18], who found that the corresponding
  asymptotic scaling form for two dimensional Potts models was
  only reached for systems of linear dimensions $L$ as big as
  5 to  10 times the correlation length $\xi$ of the infinite system,
  it seems highly desirable to check more finite-size scaling
  predictions for the model considered above.
  This will be done in Section 4.

   As we will see, the asymptotic FSS-form obtained from equation
  (1.1) will again only be reached for fairly large volumes.
  While these volumes are easily accessible for $\beta=0.5$
  and $\beta=0.47$ (the corresponding correlation length
  $\xi$ is $2.19238...$ and $4.35091...$), it is not accessible for
  temperatures near $T_c$ (we did do a third simulation for
  $\beta=.45$, corresponding to $\xi=13.5096...$). For these
  temperatures, however, a scaling ansatz using the length
  scale $\xi$ to rescale all lengths in consideration,
  should be reasonable. In fact, we will see in Section 5,
  that this ansatz allows to fit all data onto a unique curve
  which describes the crossover between first and second
  order phase transitions.

\vfill\eject

\noindent
{\bf 2. The Model and its Relation to the standard Ising Model}
\medskip

   In the next three sections, we will analyse
   the FSS for the model (1.2). We recall that
  the low temperature phases of the model consist
  of a phase ``minus" corresponding to spin configurations where most
  spins $s_x$ are equal to $-1$ with small islands of $0$ or $+1$,
   and a phase ``plus" corresponding to spin configurations which are
  a mixture of $0$ and $+1$ with small islands of $-1$.
  As we will see below, it can be shown that for $\beta>\beta_c$,
  where $\beta_c$ is the critical temperature of the Ising model,
  the phase minus is stable as long as $h$ is smaller
   then $h_t=-\beta^{-1}{\rm arsinh(1/2)}$,
  while the plus phase is stable for $h>h_t$.
   At $h=h_t$, the infinite volume magnetisation jumps from a value
  $m_-=m_-(\beta)$ to a value $m_+=m_+(\beta)>m_-$.

   In order to compare the predictions of FSS theory with
  numerical data, one needs the numerical values of the infinite volume
   magnetisation
   $m_\pm$ and susceptibility $\chi_\pm$ at the point $h=h_t\pm0$.
   As shown in [16], these constants can be obtained as follows:
   Let $Z(V,\beta,h)$ be the partition function of the
   model (1.2), and let $Z_I(V,\beta,h)$ be the partition function
   of the standard Ising model. Then $Z$ and $Z_I$ are related by the
   formula
$$
   Z(V,\beta,h)=e^{(\mu(\beta,h)-h)\beta V} Z_I(V,\beta,h)
   \eqno(2.1{\rm a})
$$
   with
$$
  \mu(\beta,h)=(1/2\beta) \, {\rm ln}({e^{\beta h}+e^{2\beta h}})
  \;.
   \eqno(2.1{\rm b})
$$

     As a consequence,
  we obtain the free energy $f(\beta,h)$ of the model (1.2) in terms
   of the free energy $f_I(\beta,\mu)$ of an Ising model with  magnetic field
  $\mu$
$$
  f(\beta,h)=f_{I}(\beta,\mu(\beta,h))+h-\mu(\beta,h)
  \;.
  \eqno(2.1{\rm c})
$$
  In a similar way, it is possible to relate the
  correlation length $\xi$ and the interface tension $\sigma$ of the
  model (1.2) at $h=h_t$ to those of the Ising model at zero field.
  One finds
$$
  \xi(\beta)=\xi_I(\beta)
  \qquad
  {\rm and}
  \qquad
  \sigma(\beta)=\sigma_I(\beta)\;.
  \eqno (2.2)
$$

  Recalling that phase transitions are associated with singularities
  of the free energy, we can read off the phase diagram of the model
  (1.2): It has the same critical temperature as the Ising model,
  while $h_t$ is just the point where $\mu(\beta,h)$ is zero, i.e.
$$
  h_t=-{1\over\beta}{\rm arsinh}(1/2)=-{0.48121183...\over\beta}
  \;.
  \eqno(2.3)
$$
   Differentiating equation~(2.1c) with respect to $h$ and taking the
  limit $h\to h_t\pm 0$, we finally obtain
  $m_\pm$ and $\chi_\pm$. In terms of the magnetisation
  $m_{I}=m_{I}(\beta)$ and the susceptibility
  $\chi_{I}=\chi_{I}(\beta)$ of the Ising model at
  $\mu=+0$ one gets
$$
  \eqalignno{
  m_\pm&=-{1\over 2}
  e^{\beta h_t}\pm {1\over 2}(2-e^{\beta h_t})m_{I}
  &(2.4{\rm a})\cr
  \chi_\pm&={1\over 2}(\beta\pm\beta m_{I})e^{3\beta h_t}
  +{1\over 4}(1+e^{2\beta h_t})^2\chi_{I}
  \;.
  &(2.4{\rm b})\cr
    }
$$
   Note that the difference $\chi_+-\chi_-$ does not involve
  the constant
  $\chi_{I}$,
$$
   \chi_+-\chi_-= (\beta m_{I})e^{3\beta h_t}\;.
   \eqno(2.4{\rm c})
$$

  We have simulated the model (1.2) in $d=2$, where
  $m_I(\beta)$ and $\xi_I=1/2\sigma_I$ are exactly known [19],
$$
  \eqalignno{
  m_I(\beta)&=(1-({\rm sinh}2\beta)^{-4})^{1/8}  &(2.5{\rm a})\cr
  \xi_I(\beta)&=(4\beta+2{\ln (\tanh}\,\beta))^{-1}
  \ \ \ {\rm for}\ \ \ \beta > \beta_c & \cr
   &=-(2\beta+{\ln(\tanh}\,\beta))^{-1}
  \ \ \ {\rm for} \ \ \  \beta <\beta_c
  \;.
  &(2.5{\rm b})\cr
 }
$$
  In order to compare FSS theory to numerical data, we will also
  need $\chi_+$ and $\chi_-$. Unfortunately, the susceptibility
  $\chi_I(\beta)$ of the Ising model is not exactly known.
  We therefore used the low-temperature series for $\chi_I(\beta)$
  found in
  [20] together with the method of Pad\'e approximants [21] to
  calculate $\chi_I(\beta)$. As in [22], we embodied the knowledge
  of the exact transition point and critical exponent in our
   approximation. The result is
  $$
  \chi_I(\beta)=4\beta u^2
  \left[
  \left(1+\sum_{n=1}^M p_n u^n\right)
  \left/\left((1-6u+u^2)\left(1+\sum_{n=1}^N q_n u^n
     \right)\right)\right. \right]^{7/4}
  \eqno(2.6)
  $$
   where $u=e^{-4\beta}$, and $p_n$, $q_n$ are given
  in table~1.  The factor $1-6u+u^2$ has a simple zero at
  $\beta =\beta_c$, ensuring that $\chi_I$ diverges as
   $(\beta-\beta_c)^{7/4}$. It is amusing to note that a Pad\'e
  approximant of the form (2.6) delivers the exact formula (2.5a)
  for $m_I$ if used on the strong coupling series for magnetisation.

  \bigskip
  \bigskip
  \vbox{\offinterlineskip
  \hrule
  \halign{&\vrule#&\strut\quad\hfil#\quad\cr
  height2pt&\omit&&\omit&&\omit&&\omit&&\omit&&\omit&\cr
  &$n$\hfil&&$p_n$\hfil&&$q_n$\hfil&&$n$\hfil&&$p_n$\hfil
  &&$q_n$\hfil&\cr
  height2pt&\omit&&\omit&&\omit&&\omit&&\omit&&\omit&\cr
  \noalign{\hrule}
  height2pt&\omit&&\omit&&\omit&&\omit&&\omit&&\omit&\cr
  & 1&& -11.422    && -9.9941   &&4&&  -0.83867  && -41.824  &\cr
  & 2&&  42.042    && 27.744    &&5&&  -4.9026   &&          &\cr
  & 3&& -49.064    && -5.1210   && &&            &&          &\cr
  height2pt&\omit&&\omit&&\omit&&\omit&&\omit&&\omit&\cr}
  \hrule}
\medskip

   \noindent{\it TABLE 1: Low temperature Pad\'e coefficients for
  the susceptibility of the Ising model.}

\vfill\eject

\noindent
{\bf 3. Equal Weight versus Equal Height}
\medskip

   In order to obtain the FSS of the order parameter distribution $P_V(s)$
  of an order parameter $s$
  (e.g. $s={ 1\over V}\sum s_x$ is the magnetisation per lattice site for
  spin models )
  in a cubic system with periodic boundary conditions and volume $V=L^d$,
  Binder and Landau approximated the probability distribution $P_V(s)$
  by a sum of two Gaussians, centered around
  the infinite volume magnetisation $m_-$ and $m_+$.
  As for the relative normalisation, in the original version [6]
  they added up two normalised Gaussians
  so that at the transition point $h_t$ each of the two coexisting
  phases contributes with the same {\it weight}.
  Following this ``equal weight"-rule [6] one gets
$$
  P_V(s)\cong {1\over N}
  \left(
   {1\over\sqrt{\chi_-}}
  e^{-V\beta{(s-m_-)^2\over 2\chi_-}}
  +
  {1\over\sqrt{\chi_+}}
  e^{-V\beta{(s-m_+)^2\over 2\chi_+}}
  \right)
  e^{\beta s(h-h_t)V}
  \;,
  \eqno(3.1{\rm a})
$$
  where $\chi_\pm$ are the infinite volume susceptibilities at the point
  $h_t-0$ and $h_t+0$, respectively, and $N=N(\beta,h)$ is chosen in such a way
  that $P_V(s)$ is normalised. Led by mean-field arguments they revised this
  version for the relative normalisation, adding up the two Gaussians
  in such a way that
  the peak {\it heights} are equal at the transition point $h_t$.
  Following this ``equal height"-rule [7] $P_V(s)$ turns out to be
$$
  P_V(s)\cong {1\over N}
  \left(
  e^{-V\beta{(s-m_-)^2\over 2\chi_-}}
  +
  e^{-V\beta{(s-m_+)^2\over 2\chi_+}}
  \right)
  e^{\beta s(h-h_t)V}
   \,.
   \eqno(3.1{\rm b})
$$
  It is an easy exercise to calculate the FSS for the finite volume
  magnetisation $m(V,h)=\int s P_V(s) ds$ following from (3.1a) and (3.1b).
   For the point $h_{\chi}(V)$ where the susceptibility
 $\chi(V,h)=dm(V,h)/dh$ has its maximum,
  a lengthy but straight-forward calculation gives
   the FSS predictions
$$
  h_{\chi}(V)=h_t+{6(\chi_+-\chi_-)\over \beta^2(m_+-m_-)^3}{1\over V^2}
  +O(V^{-3})
  \eqno(3.2{\rm a})
$$
  and
\vfill\eject
$$
  \eqalignno{
  h_{\chi}(V)&=h_t-{{{\ln}}(\chi_+/\chi_-)\over 2\beta(m_+-m_-)}{1\over V}
  \cr
  &+{\left(6-{1\over 8}{\ln}^2(\chi_+/\chi_-)    \right)}
  {\chi_+-\chi_-\over \beta^2(m_+-m_-)^3}{1\over V^2}
  +O(V^{-3})
  \;,
  &(3.2{\rm b})}
$$
   respectively.
  We emphasize that the shift predicted by the ``equal weight"-rule [6]
  is proportional to $1/V^2$
  while the shift predicted by the ``equal height"-rule [7]
  is proportional to $1/V$. (For symmetric phase transitions both rules
  are equivalent and predict that $h_{\chi}(V)=h_t$.)

\midinsert
\vskip 8.5truecm
\noindent\includegraphics{bkrfig1a.ps}
\noindent{\it \noindent Fig.~1a:
  The position $h_\chi(V)$ of the susceptibility maximum
  for $\beta=0.45$. Open points are Monte Carlo results,
   solid points are exact results (see appendix).
  The solid line is the theoretical prediction  {\rm (3.2a)},
while the dashed line is the prediction {\rm (3.2b)}.
}
\endinsert

   We have tested the predictions (3.2) for three different temperatures,
  namely for $\beta=0.45$, $\beta=0.47$ and $\beta=0.5$.
   In our simulation we first generated the probability distribution
   $P_V(s)$ for $L=4,5,...,12,14,18,30$ at $h=h_t$ using
   a multimagnetical version [23] of the
   heat bath algorithm.
   Applying the reweighting technique
  of Ferrenberg and Swendsen [24],
   we obtain the probability distribution $P_V(s)$ and hence
  the magnetisation, $m(V,h)=\sum s P_V(s)$
  and the susceptibility,
  $\chi(V,h)=\beta V\{\sum s^2P_V(s) - (\sum s P_V(s) )^2\}$,
  for $h$ in the vicinity of $h_t$. In order to use the
  vector structure of the CRAY X-MP and Y-MP, we calculated 64 independent
   copies of the system in parallel, with 15 million sweeps for each of them.
   Grouping 8 of them together to save memory, we obtained 8 statistically
   independent data sets for each $L$, and therefore 8 statistically
   independent estimates for $h_\chi(V)$.
   Statistical error estimates are obtained from these data in the
   standard way.

\midinsert
\vskip 8.5truecm
\noindent\includegraphics{bkrfig1b.ps}
\noindent{\it Fig.~1b:
  The position $h_\chi(V)$ of the susceptibility maximum
  for $\beta=0.47$.
}
\endinsert

\midinsert
\vskip 8.5truecm
\noindent\includegraphics{bkrfig1c.ps}
\noindent{\it Fig.~1c:
   The position $h_\chi(V)$ of the susceptibility maximum
   for $\beta=0.50$.
}
\endinsert

  We have plotted the theoretical curves
  (3.2a) and (3.2b) together with our numerical data in Figs.~1a-c.
  We recall that the coefficient in (3.2a) is known exactly, while
  the coefficients in (3.2b) have been obtained with the help of a
  Pad\'e improved low-temperature expansion.
  Within the resolution of Fig.~1,
  the errors
  for the numerically determined points $h_\chi(V)$
  are invisible, except for $\beta=0.45$.
  Fig.~1 shows overwhelming evidence for the equal weight
  prescription.

  Convinced of the ``equal weight"-rule we tried
  the new method to determine the
  transition point $h_t$ of first-order phase transitions
  recently proposed by C. Borgs and W. Janke [25].
  This method just {\it uses} the fact that
  both phases should have equal weight at $h=h_t$.
  To be precise, one {\it defines} a finite volume transition point
 $h_{EW}(V)$ as the point where
$$
 \sum_{s<s_{min}} P_V(s)=\sum_{s>s_{min}}P_V(s)
 \;.
 \eqno(3.3)
$$
   Here $s_{min}$,
  $s_-<s_{min}<s_+$, is the point $s$  where the distribution $P_V(s)$
  has its minimum. According to Ref.~[25], it is expected that
  the systematic errors $|h_{EW}(V)-h_t|$ are exponentially small. Here
   they are so small that $h_{EW}(V)$ agrees with $h_t$
   within the statistical errors as soon as $L/\xi$ is noticeably
   bigger then one (see Table~2 at the end of this paper).

\vfill\eject

\noindent
{\bf 4. Asymptotic First-Order FSS-Scaling}
\medskip

  In this section we test the FSS predictions following from
  (1.1)
   by Taylor expanding $f_\pm(\beta,h)$ around $h=h_t$,
$$
  f_\pm(\beta,h)
  =f(\beta,h_t) - m_\pm(h-h_t)
  -{\chi_\pm\over 2}(h-h_t)^2
  +{\chi_\pm^\prime\over 3!}(h-h_t)^3
  + \cdots
  \;.
\eqno(4.1)
$$
   For the maximum $\chi_{\rm max}(V)$ of the susceptibility,
  its position $h_\chi(V)$ and the magnetisation
  $m(h_\chi(V))$ at the point $h_\chi(V)$, the
  FSS predictions following from (1.1) are
$$
  h_\chi(V)
  =h_t
 + {6(\chi_+-\chi_+)\over \beta^2(m_+-m_-)^3}{1\over V^2}
 + O\big({1\over V^3}\big)
\eqno (4.2)
$$
$$
  m(h_\chi(V))
  ={m_++m_-\over 2}
 + {3\over 2\beta}{\chi_--\chi_+\over m_--m_+}{1\over V}
 + O\big({1\over V^2}\big)
\eqno(4.3)
$$
$$
  \chi_{\rm max}(V)
  =\beta\left({m_+-m_-\over 2}\right)^2 V
 + {\chi_++\chi_-\over 2}
 + O\big({1\over V}\big)
\;.
\eqno(4.4)
$$
   It is also interesting to analyse the FSS of the susceptibility
  and magnetisation at the infinite volume transition point $h_t$
  (for systems where $h_t$ is not known exactly, one could use the
  point $h_{EW}(V)$ introduced in [25]). Due to the exponentially
  small error bound in (1.1), the corrections to the known leading
  order FSS predictions are exponentially small
$$
m(h_t)
={m_++m_-\over 2}
\left(1+O(Ve^{-L/L_0})\right)
\eqno(4.5)
$$
$$
\chi(h_t)
=\left[\beta\left({m_+-m_-\over 2}\right)^2 V
 + {\chi_++\chi_-\over 2}\right]
\left(1+O(Ve^{-L/L_0})\right)
\;.
\eqno(4.6)
$$
  Note that the error bounds in (4.5) and (4.6) are only {\it bounds}.
  For the model considered here, the value of
  $m(h_t)$, e.g., is in fact exactly $(m_+-m_-)/2$,
  due to (2.1) and the fact that
  the finite volume magnetisation with periodic boundary conditions
  and no external field is zero for the standard Ising model.
  The corrections for $\chi(h_t)$, on the other hand, are expected to be
  asymptotically proportional to $L^\alpha e^{-L/\xi}$, with a constant
  $\alpha$ not necessary equal to the worst case bound (4.6).

  In addition to the FSS of the susceptibility and magnetisation,
  we also analyse the FSS of the cumulant $U_V$ introduced by
  Binder [5]. It is defined as
$$
U_V=
 - {
	\left\langle M^4\right\rangle _{c}
  \over
	3 \left\langle M^2\right\rangle _{c}^2
  }
   =
 - {
	\left\langle (M-\langle M\rangle )^4\right\rangle
	-  3 \left\langle (M-\langle M\rangle )^2\right\rangle ^2
   \over
	3 \left\langle (M-\langle M\rangle )^2\right\rangle ^2
   }
\eqno(4.7)
$$
  Here $\langle \cdot\rangle _{c}$ denotes connected expectation values and
  $M=\sum_{x\in V}s_x$. By the inequality
  $\langle F^2\rangle \geq\langle F\rangle ^2$
  (applied to $F=(M-\langle M\rangle )^2$) the cumulant $U_V$ cannot
  exceed the value $2/3$.
 In the high temperature region, where the asymptotic
  probability distribution of $M$ is gaussian, $U_L$ goes to
  $0$ in the infinite volume limit for all $h$,
  while in the low temperature region considered here
$$
  U_V(h_t)={2\over3}\left[1-{4(\chi_++\chi_-)\over\beta (m_+-m_-)^2}{1\over V}
  +O({1\over V^2})\right]
  \;,
  \eqno(4.8)
$$
  implying that $U_V$ goes to $2/3$ at the first order transition point
  $h=h_t$ if $V\to\infty$. At the second order transition point
  $h=h_t$, $\beta=\beta_c$ the cumulant
  $U_V$ is expected to go
  to a non-zero limit strictly smaller then $2/3$.

   In most theories $h_t$ is not known. Instead of $U_V(h_t)$,
  it is therefore more natural to consider the maximum
  $U_V^{\rm max}$ of $U_V$ as a function of $h$, and its position
  $h_U(V)$. Again, the FSS of these two quantities can be obtained
  from (1.1). One obtains
$$
  h_U(V)=h_t
 + {4(\chi_+-\chi_+)\over \beta^2(m_+-m_-)^3}{1\over V^2}
 + O\big({1\over V^3}\big)
\eqno (4.9)
$$
$$
U_V^{\rm max}=
{2\over3}
\left[
1 - {4(\chi_++\chi_-)\over\beta (m_+-m_-)^2}{1\over V} + O({1\over V^2})
\right]
\;,
\eqno(4.10)
$$

  Even though (1.1) is rigorous for large $\beta$, and probably correct
  for all $\beta>\beta_c$ provided $V$ is large enough, it is not clear
   where the large $V$ asymptotics for the first-order FSS
   predictions (4.2) through (4.10) sets in. In fact, warned by the
  results of Billoire and co-workers [18], we expect that the
  asymptotic FSS sets only in once $L\gg\xi$, where $\gg$ might mean
  five to ten times larger.

  We have tested the above finite-size scaling predictions
  for three different temperatures,
  namely for $\beta=0.45$, $\beta=0.47$ and $\beta=0.5$
  (corresponding to $\xi=13.5096...$, $\xi=4.35091...$
  and $\xi=2.19238...$).
   In our simulation we measured the quantities
  $h_\chi(V)$, $m(h_\chi(V))$, $\chi_{\rm max}(V)$, $\chi(h_t)$,
  $h_U(V)$, $U_V^{\rm max}$ and $U_V(h_t)$
  for $L=4,5,...,12,14,18,30$, see Section 3 above for the details
  of the simulation. The results are given in Table~2a-c.

  We show both the theoretical prediction according to (4.2) through
  (4.10) and our numerical results for
  $h_\chi(V)$, $h_U(V)$, $m(h_\chi(V))$, $\chi(h_t)$
  and $U_V(h_t)$ in Figs.~2~-~5 below. Monte Carlo data are represented by
  open circles and open squares, and theoretical predictions are
   represented by full or dashed lines.

   As a first point we want to
   stress that our  theoretical calculations predict that
  both $h_\chi(V)-h_t$ and $h_U(V)-h_t$ go to zero like
  $1/V^2$, a peculiar feature of an asymmetric first order
  transition with two coexisting phases\footnote{$^1$}{More generally,
  $h_\chi(V)-h_t\sim 1/V^2$ and $h_U(V)-h_t\sim 1/V^2$ if the number
  $N_-$ of phases stable for $h<h_t$ and the number $N_+$
  of phases stable for $h>h_t$ are equal. If $N_+ \neq N_-$ the
  shift in $h$ is proportional to $1/V$.   }. Indeed, our
  numerical data conform with this prediction, see Fig.~2a.
  Since the last points in Fig.~2a
  are barely distinguishable, we show a blow up of this figure
  in Fig.~2b. Unfortunately, the statistical errors for the corresponding
  data points are so big that our data for
  $h_\chi(V)$ and $h_U(V)$ agree for the last five lattice sizes
  $L=$11, ..., 30, see Table~2. We therefore decided to
   present data points  from an exact transfer matrix
  calculation in Fig.~2b. As can be seen from the figure, these
  data are in perfect agreement with the theoretical
  predictions (4.2) and (4.9).

\midinsert
\vskip \vsize
\vskip -1.8truecm
\noindent\includegraphics{bkrfig2a.ps}
\noindent{\it Fig.~2a:
The finite-size transition points
$h_\chi(V)$ (dashed line) and $h_U(V)$ (solid line).
}
\endinsert

\midinsert
\vskip \vsize
\vskip -.8truecm
\noindent\includegraphics{bkrfig2b.ps}
\noindent{\it Fig.~2b:
$h_\chi(V)$ and $h_U(V)$
(blow-up).
}
\endinsert

\midinsert
\vskip \vsize
\vskip -.8truecm
\noindent\includegraphics{bkrfig3.ps}
\noindent{\it Fig.~3:
The magnetisation at the susceptibility peak $m(h_\chi)$.
}
\endinsert

\midinsert
\vskip \vsize
\vskip -.8truecm
\noindent\includegraphics{bkrfig4.ps}
\noindent{\it Fig.~4:
The susceptibility $\chi(h_t)$
 and its asymptotic behaviour.
}
\endinsert

\midinsert
\vskip \vsize
\vskip -.8truecm
\noindent\includegraphics{bkrfig5.ps}
\noindent{\it Fig.~5:
The cumulant $U_V(h_t)$
 and its asymptotic behaviour.
}
\endinsert

  Next, we point out that the volumes for which the
  asymptotic FSS according to (4.2) through
  (4.10) sets in depends strongly on the considered quantity.
  While the asymptotic regime corresponds to
  $L/\xi\geq 1\,-\,2$ for
  $h_\xi(V)$ and $h_U(V)$, and to
  $L/\xi\geq 2\,-\,3$ for $m(h_\chi(V))$,
  both $\chi(h_t)$ and $U_V(h_t)$ require much larger volume
  ($L/\xi\geq 4\,-\,5$ and $L/\xi\geq 6\,-\,8$, respectively).
  This conforms to the results of Gupta and Irb\"ack, who
  observed similar effects for a symmetric field driven
  transition [26].

  A last point of interest is related to the determination
  of the transition point $h_t$ as
  discussed at the end of the last section [25,27].
  In addition to the point $h_{EW}(V)$ defined in (3.3),
  one may consider ``single peak" magnetisations and
  susceptibilities $m_\pm(V)$ and $\chi_\pm(V)$ defined as
$$
  m_\pm(V)=\sum s P_V^\pm(s)
\;
\eqno(4.11)
$$
  and
$$
  \chi_\pm(V)=\beta V\left(\sum s^2P_V^\pm(s)
           - \left(\sum s P_V^\pm(s) \right)^2\right)
  \;,
\eqno(4.12)
$$
  where $P_V^\pm$ is the normalised probability distribution
  of the left and right peak, respectively. More explicitly,
$$
\eqalign{
m_-(V)&=\sum_{s<s_{min}} s P_V(s)/\sum_{s<s_{min}}  P_V(s)\cr
m_+(V)&=\sum_{s>s_{min}} s P_V(s)/\sum_{s>s_{min}}  P_V(s)\cr
}
\eqno(4.13)
$$
  and similarly for $\chi_\pm(V)$.

  The corresponding data can be found in Table~2a-c as well.
  An inspection of these data for $\beta=0.50$ shows an interesting
  feature. For small system sizes,
  $\chi_\pm(V)$ are smaller then the infinite volume values
  $\chi_\pm$, and $m_+(V)>m_+$ and while $m_-(V)<m_-$.
  As the system becomes larger, all these quantities approach the
  infinite volume value, and in fact overshoot, before they finally
  approach the infinite volume limit from the other side, see Fig.~6.
  We have observed the same effect for even lower
  temperatures (where our exact transfer matrix calculation
  can easily reach the asymptotic regime),
  and we expect that it occurs for all temperatures
  $\beta<\beta_c$ provided the lattice is chosen large enough.

  We think that this effect is due to the competing influence
  of two finite-size effects for the probability distribution
  $P_V(s)$. One of these effects concerns the fact that
  the single peaks $P_V^\pm$ obtained by just cutting
  the original distribution at the point $s_{min}$
  contain events which actually should be considered part of
  the other phase. A minute of reflection
  shows that this effect decreases the susceptibility,
  and shifts the magnetisations  $m_\pm(V)$ in the direction
   observed for small lattice systems
  (if one {\it assumes} that the original distribution
  is of the form (3.1a), this can be shown rigorously). The second,
  competing effect comes from tunneling events.
  These events tend to shift the magnetisations towards
  each other, and to increase the susceptibility.
  Since overlap effects between the two peaks
  die out exponentially with the volume $V=L^d$, while
  tunneling effects die out exponentially with $L^{d-1}$,
  the competition of these two effects seems to
  naturally explain the observed overshooting for
    $m_\pm(V)$ and $\chi_\pm(V)$.

\midinsert
\vskip 8.5truecm
\noindent\includegraphics{bkrfig6.ps}
\noindent{\it Fig.~6:
The volume dependence of the ``single peak" magnetisation
$m_+(V)$ for $\beta=0.5$. The dashed line shows the infinite
volume value.
}
\endinsert

\vfill\eject

\noindent
{\bf 5. The Crossover from First- to Second-Order FSS}
\medskip

     We have seen that the formulae of sections 3 and 4
  break down when the correlation length $\xi$ becomes
  comparable with the lattice size $L$. In this section
  we try to explain finite size behaviour in this large
  $\xi$ region.

     In Fig.~7 we show the susceptibility at $h_t$ as a
  function of lattice size for a range of inverse
  temperatures $\beta$ in the neighbourhood of $\beta_c$.
  The results for the smaller lattices (up to size
  $14 \times 14$) are taken from exact calculations
  of the magnetisation distribution (see appendix).
  The susceptibilities on the larger lattices are
  Monte Carlo results, either from our own simulations
  with the three-state model (1.2) or derived (via
  the formulae in section 2) from simulations of the
  two dimensional Ising model [23,29].  What we see is
  that on small lattices the curves are all rather similar,
  independent of whether $\beta$ is above, below or at the
  critical point. This is indeed reasonable, on a finite
  lattice all physical quantities  are analytic functions
  of the coupling constants so there can be no sharp distinction
  between the first order phase transition above $\beta_c$, the
  second order phase transition at $\beta_c$ or the rapid crossover
  (without a phase transition) seen below $\beta_c$.

    This suggests that we should try to understand
  this region by using the results of the finite size scaling
  scaling theory in the neighbourhood of a second order phase
  transition. (For reviews see [10,28]).  We expect this theory
  to apply whenever the correlation length $\xi$ and lattice
  size $L$ are both large compared to the lattice spacing.

     The basic hypothesis is that when the correlation length
  is large enough thermodynamic properties are no longer
  dependent on the lattice spacing itself. By this is meant
  that instead of being
  a function of the three parameters $\beta$, $h$ and $L$
  the free energy and correlation length become functions
  of only two variables, i.e.
\vfill\eject

     \midinsert
    \vskip \vsize
    \vskip -1.05truecm
    \vskip -1.3truept
    \noindent\includegraphics{bkrfig7.ps}
    \noindent{\it Fig.~7:
    $\chi(h_t)/V$ as a function of lattice size. Solid points are
      exact results, open points are Monte Carlo results.
      }
     \endinsert

  $$
\eqalignno{
   f^{\rm sing}(\beta,h,L) &\approx L^{-d} \tilde{f}( t L^{1/\nu} ,
                       H L^{(\gamma+\beta)/\nu}) &(5.1{\rm a})\cr
   {\rm and \ \ \ \ \ \ }&&\cr
   \xi (\beta,h,L) &\approx L \tilde{\xi}( t L^{1/\nu} ,
                       H L^{(\gamma+\beta)/\nu}) &{\rm (5.1b})  }
  $$
  where the thermodynamic variables are
 $t\equiv (\beta_c-\beta)/\beta$ and $H \equiv h-h_t$,
 and $\gamma$,$\beta$ and $\nu$ are the critical exponents of
 the model (1.2), which (as explained in section 2) are the
 same as those of the Ising model, namely $\gamma = 7/4$,
 $\beta=1/8$ and $\nu = 1$. (It is unfortunate that
 the traditional symbol for inverse temperature and that for
 the critical exponent are both $\beta$, we hope that the context
 will always make it clear in which sense $\beta$ is to be
 understood.)

    Differentiating the free energy leads to a scaling form
  for the magnetic susceptibility
 $$
  \chi(\beta,h,L) \approx L^{\gamma/\nu} \tilde{\chi}( t L^{1/\nu} ,
                       H L^{(\gamma+\beta)/\nu}).  \eqno(5.2)
 $$
   (The ``hyperscaling" relationship $2 \beta + \gamma = d \nu$ has
  been used to simplify this expression.)
   A simple test of the scaling form (5.2) comes from
 our data at $h=h_t$. The finite size scaling prediction is
 $$
  \chi(\beta,h_t,L)/L^{\gamma/\nu} \approx
        \tilde{\chi}( t L^{1/\nu},0),  \eqno(5.3)
 $$
 i.e. that all data should lie on a universal curve if
 $\chi(\beta,h_t,L)/L^{\gamma/\nu}$ is plotted against
 $t L^{1/\nu}$. The results are shown in Fig.~8.
 In place of $t L^{1/\nu}$ we have used the somewhat more
 physical coordinate
$$
  \eqalignno{
     (4\beta+2{\ln (\tanh}\,\beta)) L^{1/\nu}
    &=  (L/\xi(\beta,h_t,\infty))^{1/\nu}
  \ \ \ {\rm for}\ \ \ \beta > \beta_c & \cr
    &= -2 (L/\xi(\beta,h_t,\infty))^{1/\nu}
  \ \ \ {\rm for} \ \ \  \beta <\beta_c
  &.{\rm(5.4)}\cr
 }
$$
  This is
 possible in this model since we have an exact expression
 (2.5b) for the correlation length. (The factor $-2$
 in the symmetric phase is there to ensure that the
 $x$-coordinate is an analytic function of $\beta$.)
  We see that the data
  do indeed fall on a universal curve. (Some deviation
  is seen at $\beta=0.50$, which is to be expected
  since the correlation length is only $\approx 2.19$).
  The curves predicted by the ansatz (3.1a) for first
  order phase transitions are also plotted in this figure.
  We see that the predictions are reliable as long as
  $L/\xi > 4$. The asymptotic results calculated in
  section 3 allow us to say more than the scaling
  hypothesis (5.1), because they give us the large
  $x$ asymptote of the scaling function $\tilde{\chi}(x,0)$.

\midinsert
\vskip 8.5truecm
\noindent\includegraphics{bkrfig8.ps}
\noindent{\it \noindent Fig.~8:
   A scaling plot of $\chi(h_t)$.
   Exact results are shown as filled
  symbols, Monte Carlo results as open. Only data for $L \ge 3$
  are plotted. The curves show the expected large volume behaviour.
  The dashed curves are the large volume asymptotes for
  $\beta=0.30,0.47$ and $0.50$. The solid curves are the large
  volume asymptotes at $\beta = \beta_c \pm \epsilon $.
 }
\endinsert

     The scaling form (5.2) also has implications for $h_\chi$,
  the position of the susceptibility maximum.
   $$
  h_\chi(\beta,L)-h_t(\beta) \approx
      L^{-(\gamma+\beta)/\nu} \tilde{H} ( t L^{1/\nu} )   \eqno(5.5)  $$

   Taking higher derivatives of the free energy leads to scaling
  expressions for any cumulant of the magnetisation distribution,
  in particular to a scaling expression for the Binder parameter
  $U_V$
 $$
  U_V(\beta,h,L) \approx \tilde{U}( t L^{1/\nu} ,
                       H L^{(\gamma+\beta)/\nu})  \eqno(5.6{\rm a})
 $$
  and its maximal value
    $$
  U_V^{\rm max} \approx \tilde{U}^{\rm max}(t L^{1/\nu}).
       \eqno(5.6{\rm b})
  $$
  The magnetic field at which $U_V$ reaches its maximum is given
  by a relationship of the form (5.5).

  The scaling predicted in (5.6b) is tested in Fig.~9.
  As in section 4 we find that larger
  lattices are needed to see finite size scaling for $U_V$ than for
  $\chi$.  Scaling only works well on lattices with
  $L$ larger than about 5, on smaller lattices the value of
  $U_V^{\rm max}$ lies noticeably below the scaling curve.

\midinsert
\vskip 8.5truecm
\noindent\includegraphics{bkrfig9.ps}
\noindent{\it \noindent Fig.~9:
   A scaling plot of $U_V^{\rm max}$.
   Exact results are shown as filled
  symbols, Monte Carlo results as open. Only data for $L \ge 5$
  are plotted. The curve is the large volume asymptote (4.10)
 at $\beta=\beta_c+\epsilon$
}
\endinsert

   In conclusion we see that there is a considerable region of overlap
  between the domain of validity of the first order finite size scaling
  relationships presented in earlier sections and the scaling region
  results presented here. The first order results require that $L$
  be large with respect to the correlation length, but make no
  demands on what the correlation length must be, while the results
  of this section require that $L$ and $\xi$ are both large, but
  do not restrict the ratio $L/\xi$. The domain of mutual validity
  is therefore $L \gg \xi \gg 1$, while in the entire range
  $L \gg 1$ we can always apply at least one of the formulae.

\vfill\eject
\noindent

{\bf Appendix - Exact Calculations }
\medskip

   Naively one might think that exact calculations of the partition
  function for our model would be completely impractical, since the
  number of configurations that need to be considered is $3^{L^d}$.
  However by carefully dividing the lattice up into smaller
  sub-lattices it is possible to construct algorithms in which the
  amount of calculation grows exponentially with $L^{d-1}$ rather
  than $L^d$, which gives spectacular savings in time, especially
  when $d=2$.

     The essential insight is that the partition function for a lattice
   spin model can be calculated recursively, see [30,31].

\midinsert
    \vskip \vsize
    \vskip -.8truecm
\noindent\includegraphics{bkrfig10.ps}
    \vskip -.8truecm
    \vskip -11.5truept
\noindent{\it \noindent Fig.~10:
  The procedure for adding a spin to the lattice when
  calculating the partition function recursively. (See
  the text for a detailed explanation.)
}
\endinsert

     Suppose that the partition function $Z_N$ for the $N$ site
   sub-lattice
   shown at the top of Fig.~10 is a known function of the spins on
  the boundary, ($\sigma_1$ to $\sigma_{2L}$). The spins
  $\tilde{\sigma}$ in the
  interior of the lattice (the gray region) have been summed over.
  Periodic boundary conditions in the $x$ direction are to be understood.
   $$ Z_N(\sigma_1,\cdots,\sigma_L;\sigma_{L+1},\cdots,\sigma_{j-1};
  \sigma_j,\cdots,\sigma_{2L})  $$
 $$  \equiv \sum_{\tilde{\sigma}} {\rm exp} \left( -\beta
    H_N(\tilde{\sigma},\sigma_1,\cdots,\sigma_{2L}) \right)  \eqno(A.1)
  $$

   The lattice size is next increased by one by adding the new
  spin $\sigma'_j$. The energy difference, $\delta H$, between the
  new lattice and the old lattice depends only on the boundary spins,
  not on any of the interior spins (in the example shown the energy
  change depends only on the three spins $\sigma_{j-1}, \sigma_j$ and
  $\sigma'_j$, and it is also true for the special cases of the first
  and last spin in a row that only boundary spins are involved in
  the energy change).

     Finally we can find the partition function for the larger lattice
   by summing over $\sigma_j$. Algebraically
   $$ Z_{N+1}(\sigma_1,\cdots,\sigma_L;
   \sigma_{L+1},\cdots,\sigma'_j;
  \sigma_{j+1},\cdots,\sigma_{2L})  $$
 $$   = \sum_{\sigma_j} {\rm exp} \left(
     -\beta \ \delta H(\sigma_{L+1},\cdots,\sigma_L; \sigma'_j) \right)$$
  $$ \times  Z_N(\sigma_1,\cdots,\sigma_L;\sigma_{L+1},\cdots,\sigma_{j-1};
  \sigma_j,\cdots,\sigma_{2L})  \eqno(A.2)
  $$

   The recursion can be started from the fact that $Z_0$ (the partition
  function of a lattice with no spins) is $1$.

    In our implementation of this procedure the amount of calculation
  grows as $q^{2L}$ and the amount of storage needed as $q^L$ for a
  q-state spin model.  The basic reason that the calculational burden
  grows like $q^{2L}$ is that the partial partition function $Z_N$
  depends on the $2L$ boundary spins, which means that its arguments
  can take $q^{2L}$ different values. The number of arguments would
  be  greatly
  reduced if the boundary conditions in the $y$ direction allow
  us to sum over the spins in
  the first row (the spins $\sigma_1$ to $\sigma_L$). Examples of boundary
  conditions which permit such a sum are either free boundary
  conditions or a fixed boundary configuration. These would reduce the
  computational time to a value $\propto q^L$.  Even though they are
  not the best boundary conditions for exact calculations we have used
  periodic boundary conditions in both directions to make comparison
  with Monte Carlo results possible.

     The close relationship of our model to the Ising model means
  that we can calculate partition functions for the Ising model
  which has $q=2$ and then transform to the $q=3$ model at the
  end, meaning that computation grows as $4^L$ instead of $9^L$.

    Even though the calculation grows exponentially we were able
  to handle surprisingly large lattices, with sizes of up to
  $14 \times 14$, without needing a supercomputer.

    Once we have found the multiplicity of states $N_I(E,n_-)$ for
 the Ising model it is
 a simple matter to use the binomial distribution to calculate the
 corresponding multiplicities  for our model (1.2)
  $$
   N(E,n_-,n_0) = N_I(E,n_-) {(L^d-n_-)! \over n_0! (L^d-n_- -n_0)!}
  \eqno(A.3)
  $$
  Here $E$ is the number of broken bonds and $n_-, n_0$ and $n_+$
  are the number of spins of a given type. We have a simple relationship
  of this type because boundaries between spin $0$ and spin $+1$
  entail no energy cost while the energy cost for boundaries between
  spin $-1$ and either of the other two spin types is the same in both
  cases.  It is a simple matter to derive relationships between
  partition functions and magnetisation histograms of the two
  models by summing (A.3)
  over (some of) its arguments.

\vfill\eject

\noindent
{\bf References}
\baselineskip=15pt
\medskip

\item{[1]}  M.E. Fisher, in {\it  Critical Phenomena},
Proceedings of the Enrico Fermi International School of
Physics, Vol. 51, M.S. Green, ed. (Academic Press, New York, 1971).

\item{[2]}  M.E. Fisher,  M.N. Barber,
{\it Phys. Rev. Lett.} {\bf 28} (1972) 1516.

\item{[3]}  M.E. Fisher and  A.N. Berker,
{\it Phys. Rev.} {\bf B26} (1982) 2507.

\item{[4]} H.W.J. Bl\"ote, M.P. Nightingale,
{\it Physica} {\bf 112A} (1981) 405.

\item{[5]} K. Binder, {\it Z. Phys.} {\bf B43} (1981) 119.

\item{[6]}  K. Binder, D.P. Landau,
{\it Phys. Rev.} {\bf B30} (1984) 1477.

\item{[7]}  M.S.S. Challa, D.P. Landau, K. Binder,
{\it Phys. Rev.} {\bf B34} (1986) 1841.

\item{[8]} V. Privman, M.E. Fisher,
{\it J.\ Stat.\ Phys.} {\bf 33} (1983) 385.

\item{[9]} V. Privman, in {\it Finite-Size Scaling and Numerical
Simulation of Statistical Systems},
ed. V. Privman, (Singapore, World Scientific, 1990).

\item{[10]} K. Binder ``Finite Size Effects at Phase Transitions",
in {\it Computational Methods in Field Theory},
(Schladming, Feb 1-8, 1992).

\item{[11]} C. Borgs, R. Koteck\'y,
{\it J.\ Stat.\ Phys.} {\bf 61} (1990) 79 and
{\it Phys.\ Rev.\ Lett.} {\bf 68} (1992) 1734.

\item{[12]} C. Borgs, R. Koteck\'y, S. Miracle-Sole,
{\it J.\ Stat.\ Phys.} {\bf 62} (1991) 529.

\item{[13]} S. Pirogov and Ya. Sinai,
{\it Theor. Math. Phys.} {\bf 25} (1975) 1185
and {\bf 26} (1976) 39.

\item{[14]} As a phenomenological ansatz, a formula of the form (1.1)
  was used by several authors, see
  e.g. [9,15] for the description of the underlying ideas.
  At  $h=h_t$, this formula has been rigorously proven
  for a wide class of low temperature systems in
  C. Borgs, J. Imbrie, {\it Commun. Math. Phys.}
  {\bf 123} (1989) 305. In ref. [11], the analysis of Borgs and Imbrie
   was generalised to $h\neq h_t$, thus establishing a rigorous basis
  for a finite-size scaling theory based on equation~(1.1).

\item{[15]}
V. Privman, J. Rudnick, {\it J. Stat. Phys.} {\bf 60} (1990) 551.

\item{[16]} C. Borgs, S. Kappler,
{\it Phys. Lett.} {\bf A 171} (1992) 37.

\item{[17]} M. Blume, V.J. Emery, R.B. Griffiths;
{\it Phys. Rev.} {\bf A 4} (1971) 1071.

\item{[18]} A. Billoire, R. Lacaze, A. Morel, {\it Nucl. Phys.}
{\bf B370} (1992) 773;
A. Billoire, T. Neuhaus, B. Berg:
Observation of FSS for a first order phase transition.
Preprint SPhT-92/120.

\item{[19]} C.N. Yang, {\it Phys. Rev.} {\bf 85} (1952) 808.

\item{[20]} C. Domb, in
{\it Phase Transitions and Critical Phenomena}, Vol.~3,
ed. C. Domb and M.S. Green (Academic Press, London 1974)

\item{[21]} G.A. Baker,
{\it Essentials of Pad\'e Approximants},
     (Academic Press, New York, 1975)

\item{[22]} J.W. Essam, M.E. Fisher, {\it J. Chem. Phys.}
{\bf 38} (1963) 802; H.B. Tarko, M.E. Fisher,
{\it Phys. Rev.} {\bf B11} (1975) 1217.

\item{[23]} B. Berg, U. Hansmann, T. Neuhaus,
preprint, SCRI-91-125, 1992, submitted to Phys. Rev. B.

\item{[24]} A. Ferrenberg, R.H. Swendsen, Phys. Rev. Lett. {\bf 61}
(1988) 2635.

\item{[25]} C. Borgs, W. Janke,
{\it Phys. Rev. Lett.} {\bf 68} (1992) 1738.

\item{[26]} S. Gupta, A. Irb\"ack, M. Ohlsson,
 ``Finite-size Scaling at Phase Coexistence",
HLRZ preprint 27/93, April 1993.

\item{[27]} W. Janke, ``Accurate First-Order Transition Points
from Finite-Size Data without Power-Law Corrections",
HLRZ preprint 50/92, 1992.

\item{[28]} K. Binder ``Phase Transitions in Reduced Geometry",
{\it Annu. Rev. Phys. Chem. 1992. 43:33-59}

\item{[29]} U. Wolff,
{\it Phys. Rev. Lett. 62 (1989) 361}

\item{[30]} K. Binder,
{\it Physica 62 (1972) 508}

\item{[31]} G. Bhanot,
{\it J. Stat. Phys. 60 (1990) 55}

\vfill\eject

\vbox{\offinterlineskip
\hrule
\halign{&\vrule#&\strut\quad\hfil#\quad\cr
height2pt&\omit&&\omit&&\omit&&\omit&&\omit&&\omit&\cr
&L\hfil&&$h_{\chi}(V)$\hfil&&$m(h_\chi(V))$\hfil&&$\chi_{\rm max}(V)/V$
\hfil&&$\chi(h_t)/V$\hfil&\cr
height2pt&\omit&&\omit&&\omit&&\omit&&\omit&\cr
\noalign{\hrule}
height2pt&\omit&&\omit&&\omit&&\omit&&\omit&\cr
& 4&&-1.062853(44)&&-.291383(07)&&.171484(05)&&.171355(06)&\cr
& 5&&-1.066617(38)&&-.297833(08)&&.163352(07)&&.163301(06)&\cr
& 6&&-1.067975(22)&&-.301295(12)&&.157308(08)&&.157282(08)&\cr
& 7&&-1.068570(22)&&-.303355(08)&&.152602(05)&&.152587(05)&\cr
& 8&&-1.068925(25)&&-.304703(07)&&.148814(07)&&.148807(06)&\cr
& 9&&-1.069085(17)&&-.305617(12)&&.145699(07)&&.145694(07)&\cr
&10&&-1.069174(15)&&-.306261(10)&&.143083(08)&&.143080(09)&\cr
&11&&-1.069264(21)&&-.306770(11)&&.140857(06)&&.140856(06)&\cr
&12&&-1.069300(20)&&-.307118(08)&&.138951(12)&&.138951(12)&\cr
&14&&-1.069340(17)&&-.307640(11)&&.135837(08)&&.135837(08)&\cr
&18&&-1.069338(11)&&-.308173(09)&&.131496(11)&&.131495(11)&\cr
&30&&-1.069360(13)&&-.308728(30)&&.125416(17)&&.125416(18)&\cr
height2pt&\omit&&\omit&&\omit&&\omit&&\omit&\cr}
\hrule}

\vbox{\offinterlineskip
\hrule
\halign{&\vrule#&\strut\quad\hfil#\quad\cr
height2pt&\omit&&\omit&&\omit&&\omit&\cr
&L\hfil&&$h_{U}(V)$\hfil&&$U_V^{\rm max}$\hfil&&$U_V(h_t)$\hfil&\cr
height2pt&\omit&&\omit&&\omit&&\omit&\cr
\noalign{\hrule}
height2pt&\omit&&\omit&&\omit&&\omit&\cr
& 4&&-1.065260(44)&&.598796(07)&&.598403(12)&\cr
& 5&&-1.067640(38)&&.606576(08)&&.606414(08)&\cr
& 6&&-1.068483(22)&&.611434(11)&&.611349(12)&\cr
& 7&&-1.068851(22)&&.614893(07)&&.614841(10)&\cr
& 8&&-1.069093(25)&&.617588(09)&&.617564(10)&\cr
& 9&&-1.069191(16)&&.619833(10)&&.619818(11)&\cr
&10&&-1.069244(15)&&.621759(16)&&.621748(18)&\cr
&11&&-1.069313(21)&&.623476(09)&&.623473(10)&\cr
&12&&-1.069335(20)&&.625052(16)&&.625051(16)&\cr
&14&&-1.069359(17)&&.627860(11)&&.627860(10)&\cr
&18&&-1.069345(11)&&.632589(14)&&.632587(13)&\cr
&30&&-1.069361(13)&&.642953(22)&&.642953(36)&\cr
height2pt&\omit&&\omit&&\omit&&\omit&\cr}
\hrule}

\vbox{\offinterlineskip
\hrule
\halign{&\vrule#&\strut\quad\hfil#\quad\cr
height2pt&\omit&&\omit&&\omit&&\omit&&\omit&&\omit&\cr
&L\hfil&&$h_{EW}(V)$\hfil&&$m_+(V)$\hfil&&$m_-(V)$\hfil&&$\chi_+(V)$\hfil
&&$\chi_-(V)$\hfil&\cr
height2pt&\omit&&\omit&&\omit&&\omit&&\omit&&\omit&\cr
\noalign{\hrule}
height2pt&\omit&&\omit&&\omit&&\omit&&\omit&&\omit&\cr
& 4&&-1.069540(42)&&.28434(02)&&-.90380(02)&&.2488(01)&&.1523(01)&\cr
& 5&&-1.069546(38)&&.27217(02)&&-.89176(02)&&.3214(01)&&.2229(01)&\cr
& 6&&-1.069332(24)&&.26350(03)&&-.88146(03)&&.3967(02)&&.3093(01)&\cr
& 7&&-1.069321(34)&&.25587(21)&&-.87381(21)&&.4849(24)&&.3989(23)&\cr
& 8&&-1.069398(26)&&.24938(06)&&-.86779(08)&&.5832(09)&&.4918(09)&\cr
& 9&&-1.069435(27)&&.24364(22)&&-.86313(24)&&.6948(42)&&.5828(43)&\cr
&10&&-1.069362(18)&&.24004(18)&&-.85797(16)&&.7860(41)&&.7035(38)&\cr
&11&&-1.069407(22)&&.23605(10)&&-.85447(09)&&.9016(23)&&.8081(24)&\cr
&12&&-1.069402(26)&&.23292(14)&&-.85123(17)&&1.0142(48)&&.9216(50)&\cr
&14&&-1.069397(17)&&.22771(15)&&-.84606(14)&&1.2488(59)&&1.1530(61)&\cr
&18&&-1.069361(11)&&.22064(07)&&-.83897(10)&&1.7316(58)&&1.6274(66)&\cr
&30&&-1.069363(13)&&.21183(08)&&-.82994(09)&&3.036(16)&&2.944(16)&\cr
height2pt&\omit&&\omit&&\omit&&\omit&&\omit&&\omit&\cr
\noalign{\hrule}
height2pt&\omit&&\omit&&\omit&&\omit&&\omit&&\omit&\cr
&$\infty$&&-1.06935961..\hfill&&.2087522..\hfill&&-.8267862..\hfill
&&4.75\phantom{6(16)}&&4.67\phantom{6(16)}&\cr
height2pt&\omit&&\omit&&\omit&&\omit&&\omit&&\omit&\cr}
\hrule}

\medskip
\noindent{\it TABLE 2a: Monte Carlo results for $\beta=0.45$}

\vfill\eject

\vbox{\offinterlineskip
\hrule
\halign{&\vrule#&\strut\quad\hfil#\quad\cr
height2pt&\omit&&\omit&&\omit&&\omit&&\omit&&\omit&\cr
&L\hfil&&$h_{\chi}(V)$\hfil&&$m(h_\chi(V))$\hfil&&$\chi_{\rm max}(V)/V$
\hfil&&$\chi(h_t)/V$\hfil&\cr
height2pt&\omit&&\omit&&\omit&&\omit&&\omit&\cr
\noalign{\hrule}
height2pt&\omit&&\omit&&\omit&&\omit&&\omit&\cr
& 4&&-1.018099(59)&&-.2917063(77)&&.1879631(51)&&.1878343(56)&\cr
& 5&&-1.021451(34)&&-.2980674(69)&&.1817612(56)&&.1817091(46)&\cr
& 6&&-1.022640(37)&&-.3014688(70)&&.1775283(55)&&.1775017(61)&\cr
& 7&&-1.023182(23)&&-.3035004(77)&&.1744855(53)&&.1744708(53)&\cr
& 8&&-1.023500(27)&&-.3048250(81)&&.1722426(57)&&.1722357(57)&\cr
& 9&&-1.023655(16)&&-.3057188(65)&&.1705276(51)&&.1705242(51)&\cr
&10&&-1.023683(14)&&-.3063460(64)&&.1692254(71)&&.1692216(67)&\cr
&11&&-1.023747(20)&&-.3068320(83)&&.1682018(49)&&.1681995(46)&\cr
&12&&-1.023794(23)&&-.3071837(68)&&.1674027(69)&&.1674017(67)&\cr
&14&&-1.023813(14)&&-.3076732(72)&&.1662249(70)&&.1662241(71)&\cr
&18&&-1.023858(09)&&-.3082108(91)&&.1649093(29)&&.1649093(29)&\cr
&30&&-1.023846(06)&&-.3087157(58)&&.1636313(95)&&.1636305(96)&\cr
height2pt&\omit&&\omit&&\omit&&\omit&&\omit&\cr}
\hrule}

\vbox{\offinterlineskip
\hrule
\halign{&\vrule#&\strut\quad\hfil#\quad\cr
height2pt&\omit&&\omit&&\omit&&\omit&\cr
&L\hfil&&$h_{U}(V)$\hfil&&$U_V^{\rm max}$\hfil&&$U_V(h_t)$\hfil&\cr
height2pt&\omit&&\omit&&\omit&&\omit&\cr
\noalign{\hrule}
height2pt&\omit&&\omit&&\omit&&\omit&\cr
& 4&&-1.020208(59)&&.610042(05)&&.609678(15)&\cr
& 5&&-1.022328(34)&&.619780(04)&&.619626(06)&\cr
& 6&&-1.023067(37)&&.626326(06)&&.626242(11)&\cr
& 7&&-1.023413(23)&&.631209(05)&&.631161(06)&\cr
& 8&&-1.023635(27)&&.635124(05)&&.635103(08)&\cr
& 9&&-1.023740(16)&&.638356(05)&&.638348(06)&\cr
&10&&-1.023738(14)&&.641133(07)&&.641119(06)&\cr
&11&&-1.023784(20)&&.643530(05)&&.643523(05)&\cr
&12&&-1.023820(23)&&.645639(06)&&.645636(07)&\cr
&14&&-1.023828(14)&&.649127(06)&&.649124(08)&\cr
&18&&-1.023863(09)&&.654133(03)&&.654133(04)&\cr
&30&&-1.023847(06)&&.661167(06)&&.661162(09)&\cr
height2pt&\omit&&\omit&&\omit&&\omit&\cr}
\hrule}

\vbox{\offinterlineskip
\hrule
\halign{&\vrule#&\strut\quad\hfil#\quad\cr
height2pt&\omit&&\omit&&\omit&&\omit&&\omit&&\omit&\cr
&L\hfil&&$h_{EW}(V)$\hfil&&$m_+(V)$\hfil&&$m_-(V)$\hfil&&$\chi_+(V)$\hfil
&&$\chi_-(V)$\hfil&\cr
height2pt&\omit&&\omit&&\omit&&\omit&&\omit&&\omit&\cr
\noalign{\hrule}
height2pt&\omit&&\omit&&\omit&&\omit&&\omit&&\omit&\cr
& 4&&-1.024032(60)&&.303542(24)&&-.922636(09)&&.2300(01)&&.1273(01)&\cr
& 5&&-1.023969(35)&&.296404(16)&&-.915421(15)&&.2807(01)&&.1770(01)&\cr
& 6&&-1.023816(49)&&.291949(99)&&-.909913(99)&&.3280(11)&&.2320(10)&\cr
& 7&&-1.023822(22)&&.288400(97)&&-.906357(92)&&.3783(12)&&.2830(12)&\cr
& 8&&-1.023887(25)&&.285761(73)&&-.903962(76)&&.4281(11)&&.3297(12)&\cr
& 9&&-1.023896(17)&&.283968(58)&&-.902061(63)&&.4727(12)&&.3761(13)&\cr
&10&&-1.023843(15)&&.282669(71)&&-.900775(64)&&.5146(18)&&.4171(17)&\cr
&11&&-1.023858(19)&&.281679(38)&&-.899877(42)&&.5533(10)&&.4536(12)&\cr
&12&&-1.023872(23)&&.281054(34)&&-.899190(31)&&.5859(11)&&.4874(11)&\cr
&14&&-1.023857(14)&&.280179(28)&&-.898372(28)&&.6433(13)&&.5400(15)&\cr
&18&&-1.023873(09)&&.279668(27)&&-.897689(30)&&.7059(19)&&.6123(16)&\cr
&30&&-1.023848(06)&&.279645(23)&&-.897639(17)&&.7451(10)&&.6528(09)&\cr
height2pt&\omit&&\omit&&\omit&&\omit&&\omit&&\omit&\cr
\noalign{\hrule}
height2pt&\omit&&\omit&&\omit&&\omit&&\omit&&\omit&\cr
&$\infty$&&-1.02385495..\hfill&&.27967484..\hfill&&-.89770883..\hfill
&&.7378\phantom{(10)}\hfill&&.6433\phantom{(10)}\hfill&\cr
height2pt&\omit&&\omit&&\omit&&\omit&&\omit&&\omit&\cr}
\hrule}

\medskip
\noindent{\it TABLE 2b: Monte Carlo results for $\beta=0.47$}
\vfill\eject

\vbox{\offinterlineskip
\hrule
\halign{&\vrule#&\strut\quad\hfil#\quad\cr
height2pt&\omit&&\omit&&\omit&&\omit&&\omit&&\omit&\cr

&L\hfil&&$h_{\chi}(V)$\hfil&&$m(h_\chi(V))$\hfil&&$\chi_{\rm max}(V)/V$
\hfil&&$\chi(h_t)/V$\hfil&\cr
height2pt&\omit&&\omit&&\omit&&\omit&&\omit&\cr
\noalign{\hrule}
height2pt&\omit&&\omit&&\omit&&\omit&&\omit&\cr
& 4&&-.957392(41)&&-.2920229(56)&&.2112605(43)&&.2111255(57)&\cr
& 5&&-.960375(48)&&-.2982867(54)&&.2070919(30)&&.2070386(41)&\cr
& 6&&-.961410(34)&&-.3016339(73)&&.2045653(26)&&.2045385(27)&\cr
& 7&&-.961846(20)&&-.3036197(50)&&.2029542(30)&&.2029382(34)&\cr
& 8&&-.962120(32)&&-.3049161(23)&&.2018638(42)&&.2018563(46)&\cr
& 9&&-.962230(32)&&-.3057789(67)&&.2011162(70)&&.2011113(60)&\cr
&10&&-.962293(23)&&-.3064065(53)&&.2005801(45)&&.2005767(48)&\cr
&11&&-.962335(20)&&-.3068779(35)&&.2001908(63)&&.2001885(60)&\cr
&12&&-.962391(17)&&-.3072287(51)&&.1998887(67)&&.1998882(69)&\cr
&14&&-.962375(06)&&-.3076952(53)&&.1994660(22)&&.1994642(24)&\cr
&18&&-.962421(14)&&-.3082342(72)&&.1990105(36)&&.1990105(37)&\cr
&30&&-.962427(08)&&-.3087290(42)&&.1985327(29)&&.1985326(26)&\cr
height2pt&\omit&&\omit&&\omit&&\omit&&\omit&\cr}
\hrule}

\vbox{\offinterlineskip
\hrule
\halign{&\vrule#&\strut\quad\hfil#\quad\cr
height2pt&\omit&&\omit&&\omit&&\omit&\cr
&L\hfil&&$h_{U}(V)$\hfil&&$U_V^{\rm max}$\hfil&&$U_V(h_t)$\hfil&\cr
height2pt&\omit&&\omit&&\omit&&\omit&\cr
\noalign{\hrule}
height2pt&\omit&&\omit&&\omit&&\omit&\cr
& 4&&-.959187(41)&&.6220964(36)&&.6217458(11)&\cr
& 5&&-.961108(48)&&.6328211(25)&&.6326800(12)&\cr
& 6&&-.961761(34)&&.6398531(33)&&.6397789(09)&\cr
& 7&&-.962034(20)&&.6448716(16)&&.6448240(05)&\cr
& 8&&-.962230(32)&&.6486261(28)&&.6486061(09)&\cr
& 9&&-.962298(32)&&.6515315(27)&&.6515180(07)&\cr
&10&&-.962337(23)&&.6538180(23)&&.6538081(06)&\cr
&11&&-.962365(20)&&.6556541(31)&&.6556476(04)&\cr
&12&&-.962413(17)&&.6571408(20)&&.6571405(03)&\cr
&14&&-.962387(06)&&.6593655(08)&&.6593586(03)&\cr
&18&&-.962425(14)&&.6620330(08)&&.6620329(07)&\cr
&30&&-.962428(08)&&.6649091(06)&&.6649074(18)&\cr
height2pt&\omit&&\omit&&\omit&&\omit&\cr}
\hrule}

\vbox{\offinterlineskip
\hrule
\halign{&\vrule#&\strut\quad\hfil#\quad\cr
height2pt&\omit&&\omit&&\omit&&\omit&&\omit&&\omit&\cr
&L\hfil&&$h_{EW}(V)$\hfil&&$m_+(V)$\hfil&&$m_-(V)$\hfil&&$\chi_+(V)$\hfil
&&$\chi_-(V)$\hfil&\cr
height2pt&\omit&&\omit&&\omit&&\omit&&\omit&&\omit&\cr
\noalign{\hrule}
height2pt&\omit&&\omit&&\omit&&\omit&&\omit&&\omit&\cr
& 4&&-.962517(40)&&.325905(14)&&-.944591(07)&&.20538(04)&&.09385(03)&\cr
& 5&&-.962494(48)&&.323234(09)&&-.941739(09)&&.23116(02)&&.11961(03)&\cr
& 6&&-.962411(34)&&.321985(05)&&-.940006(11)&&.25043(05)&&.14281(06)&\cr
& 7&&-.962386(20)&&.321226(27)&&-.939220(26)&&.26679(35)&&.15947(34)&\cr
& 8&&-.962440(32)&&.320767(09)&&-.938869(08)&&.27967(08)&&.17092(05)&\cr
& 9&&-.962428(32)&&.320633(17)&&-.938654(18)&&.28755(27)&&.17991(31)&\cr
&10&&-.962423(23)&&.320555(16)&&-.938597(13)&&.29333(28)&&.18540(22)&\cr
&11&&-.962423(20)&&.320557(10)&&-.938592(14)&&.29611(12)&&.18939(14)&\cr
&12&&-.962454(17)&&.320545(10)&&-.938622(15)&&.29874(13)&&.19064(17)&\cr
&14&&-.962409(06)&&.320596(07)&&-.938641(07)&&.30017(11)&&.19183(09)&\cr
&18&&-.962434(14)&&.320660(09)&&-.938720(11)&&.29854(08)&&.19102(10)&\cr
&30&&-.962429(08)&&.320688(08)&&-.938717(05)&&.29677(09)&&.18925(11)&\cr
height2pt&\omit&&\omit&&\omit&&\omit&&\omit&&\omit&\cr
\noalign{\hrule}
height2pt&\omit&&\omit&&\omit&&\omit&&\omit&&\omit&\cr
&$\infty$&&-.96242365..\hfill&&.32068921..\hfill&&-.93872320\hfill
&&.29677\phantom{(09)}\hfill&&.18920\phantom{(11)}\hfill&\cr
height2pt&\omit&&\omit&&\omit&&\omit&&\omit&&\omit&\cr}
\hrule}
\medskip

\noindent{\it Table 2c: Monte Carlo results for $\beta=0.50$}
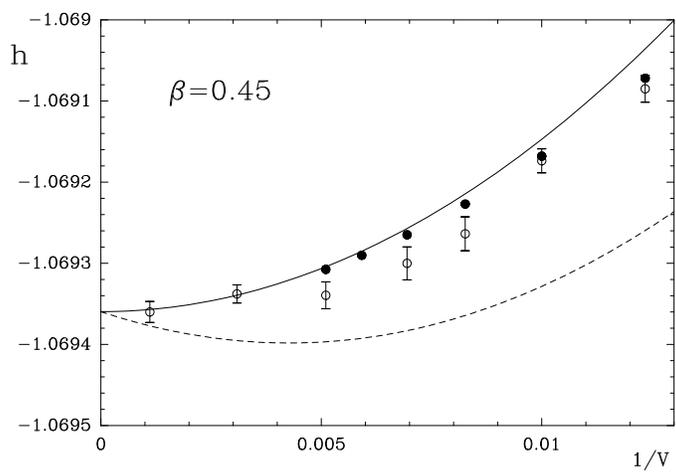
\bye